\documentclass[%
 superscriptaddress,
 preprint,
 showpacs,preprintnumbers,
 amsmath,amssymb,
 aps,
 prb,
]{revtex4-1}

\usepackage{graphicx}% Include figure files
\usepackage{dcolumn}% Align table columns on decimal point
\usepackage{bm}% bold math
\usepackage{upgreek}
\usepackage{amsmath}

\begin{document}

%\preprint{APS/123-QED}

\title{Perfect absorption in GaAs metasurfaces by degenerate critical coupling}% Force line breaks with \\

\author{Wenya Chen}
\affiliation{Institute for Advanced Study, Nanchang University, Nanchang 330031, China}
\affiliation{Jiangxi Key Laboratory for Microscale Interdisciplinary Study, Nanchang University, Nanchang 330031, China}

\author{Xing Wang}
\affiliation{Institute for Advanced Study, Nanchang University, Nanchang 330031, China}
\affiliation{Jiangxi Key Laboratory for Microscale Interdisciplinary Study, Nanchang University, Nanchang 330031, China}

\author{Junyi Duan}
\affiliation{Institute for Advanced Study, Nanchang University, Nanchang 330031, China}
\affiliation{Jiangxi Key Laboratory for Microscale Interdisciplinary Study, Nanchang University, Nanchang 330031, China}

\author{Chaobiao Zhou}
\affiliation{College of Mechanical and Electronic Engineering, Guizhou Minzu University, Guiyang 550025, China}

\author{Tingting Liu}
\email{ttliu@hue.edu.cn}
\affiliation{School of Physics and Electronics Information, Hubei University of Education, Wuhan 430205, China}

\author{Shuyuan Xiao}
\email{syxiao@ncu.edu.cn}
\affiliation{Institute for Advanced Study, Nanchang University, Nanchang 330031, China}
\affiliation{Jiangxi Key Laboratory for Microscale Interdisciplinary Study, Nanchang University, Nanchang 330031, China}

\begin{abstract}
Enhancing absorption in optically thin semiconductors is the key in the development of high-performance optical and optoelectronic devices. In this paper, we resort to the concept of degenerate critical coupling and design an ultra-thin semiconductor absorber composed of free-standing GaAs nanocylinder metasurfaces in the near infrared. The numerical results show that perfect absorption can be achieved through overlapping two Mie modes with opposite symmetry, with each mode contributing a theoretical maximum of $50\%$ in their respective critical coupling state. The absorption also shows the polarization-independent and angle-insensitive robustness. This work, together with the design concept, opens up great opportunities for the realization of high-efficiency metasurface devices, including optical emitters, modulators, detectors, and sensors. 
\end{abstract}

%\pacs{42.70.-a, 42.79.-e, 78.67.Pt}% PACS, the Physics and Astronomy
\maketitle

%\tableofcontents

\section{\label{sec1}Introduction}

Optical absorption in semiconductors has received increasing attention due to its broad application scenarios, solar cells\cite{Zhou2016, Ma2018, Wang2020a}, photodetection\cite{Tagliabue2018, Guo2020}, and biosensing\cite{Rodrigo2015, Tan2018, Ren2020, Zhao2020}, to name a few. In particular, guaranteeing high absorption in optically thin semiconductors is the key to reducing the carrier extraction time and enhancing the device performance. However, there is a ceiling of absorption ($50\%$) in all ultra thin free-standing optical film\cite{Kim2016}. To break such upper limit, a variety of strategies with the aid of metals, including the Dallenbach\cite{Dallenbach1938}, the Salisbury\cite{Salisbury1952}, and the metasurface perfect absorbers\cite{Landy2008, Liu2010, Xiong2017, Luo2018, Cheng2020} (based on period subwavelength resonant unit cells) have been proposed, in which a common used metal back mirror provides the required one-port configuration for perfect absorption, but at the cost of a broadband reflection outside the absorption band\cite{Alaee2017}. More importantly, the substantial optical absorption occurs in the metal components leads to significant photothermal conversion, namely, the Joule heat instead of photo-induced carriers is generated, which is not favored in most optical and optoelectronic devices. 

In recent years, the so-called all-dielectric meta-optics that utilizes semiconductors (Si, Ge, GaAs, etc) with the Mie resonances excited in their patterned subwavelength resonators emerges as a new milestone in the field of nanophotonics and metasurfaces\cite{Kuznetsov2016, Jahani2016, Decker2016, Koshelev2018, Li2019, Huang2021}. The high refractive index and low loss characteristics make these materials great candidates for the localization of light. Under the circumstances, achieving high absorption by directly using the Mie resonances and further overlapping these modes has been predicted in photonic crystals and metasurfaces, without the aid of metals\cite{Piper2014, Ming2017, Tian2018, Tian2020, Mitrofanov2020, Hale2020, Fan2021}. The concept of degenerate critical coupling is developed with quite strict criteria: 1) the resonator structure should support two modes of odd and even symmetries, $\omega_{1}=\omega_{2}$; 2) the radiation rate of the two modes should exactly match the dissipative loss rate of the absorbing at the resonant wavelength, $\gamma_{1}=\gamma_{2}=\delta$, and both modes will be critically coupled under such conditions, which renders the perfection absorption in a two-port configuration. As a matter of fact, such degenerate requirements are not easy to fulfill, and the perfect absorption in some of previous works indeed employs overlapping of multiple Mie resonances.

In this work, we revisit the concept of degenerate critical coupling and demonstrate perfect absorption in ultra thin semiconductor metasurfaces based on free-standing GaAs nanocylinders in the near-infrared. The proposed structure is a symmetric two-mode, two-port system with symmetry plane lying in the center of the nanocylinders perpendicular to the cylindrical axis, which is obviously different from the conventional absorbers with metal reflector mirror. Two modes, i.e., the electric dipole and magnetic dipole, possess opposite symmetries and reach their respective critical coupling state where the radiation rate is equal to the dissipative loss rate of GaAs, rendering their maximum contribution of $50\%$ at a specific radius (height) of the nanocylinders, and finally achieve the perfect absorption as a whole. The temporal coupled-mode theory (TCMT) together with rigorous multipole analysis is applied to separate the contributions of the two modes to the perfect absorption. This work lays out the foundation for the next generation high-efficiency optical and optoelectronic devices, in which the bulky semiconductors can be replaced with ultra thin perfect absorption semiconductor metasurfaces.

\section{\label{sec2}Theoretical formalism of degenerate critical coupling}

TCMT is used to derive the degenerate critical coupling conditions for perfect absorption. Here we consider the input-output behaviors of a mirror-symmetric resonator structure which supports two modes of opposite symmetries and couples with the outside through two identical ports. If there is no dissipative loss in the system, the dynamical equations can be given as\cite{Piper2014, Suh2004},
\begin{eqnarray}
\frac{da}{dt}&=&(i\omega_{0}-\gamma)a+D^{T}|s_{+}\rangle,\label{eq1} \\
|s_{-}\rangle&=&C|s_{+}\rangle+Da,\label{eq2}
\end{eqnarray}
where $a$ is a vector (rather than a number in single-mode system) which represents the resonance amplitudes, with $|a_{j}|^2$ corresponding to the energy stored in the $j$th mode. $\omega_0$ and $\gamma$ are real diagonal matrices describing the resonance frequency and radiation loss, respectively. $|s_{+}\rangle$ and $|s_{-}\rangle$ are the amplitudes of incoming and outgoing waves. $C$ is the scattering matrix of the direct process, which describes the transmission and reflection between the two ports in the absence of the resonator structure, and $D$ is the coupling matrix account for coupling between the ports and the modes, with $D^{\dagger}D=2\gamma$ and $CD^{\ast}=-D$, due to the time-reversal symmetry and energy conservation arguments. 

For the case where the incoming wave with unit amplitude is only incident from a single port, the energy stored in the resonator structure can be derived from Eqs. (\ref{eq1}) and (\ref{eq2}),
\begin{eqnarray}
|a|^{2}=\sum_{j=1}^{2}\frac{\gamma_{j}}{(\omega-\omega_{j})^2+\gamma_{j}^{2}}.\label{eq3}
\end{eqnarray}
When a dissipative loss is loaded to the above system and the basic symmetry of the resonator structure is unchanged, the loss of each mode remains independent, thus the Eq. (\ref{eq1}) can be simply modified to $da/dt=(i\omega_{0}-\gamma-\delta)a+D^{T}|s_{+}\rangle$, by adding a real diagonal matrix $\delta$, which gives the dissipative loss rate. Then the stored energy $|a|^{2}$ can be updated by amending $\gamma^{2}$ in the denominator with $(\gamma+\delta)^2$, and the absorption in the system is expressed as
\begin{eqnarray}
A=\sum_{j=1}^{2}\frac{2\delta_{j}\gamma_{j}}{(\omega-\omega_{j})^2+(\gamma_{j}+\delta_{j})^{2}}.\label{eq4}
\end{eqnarray}
It can be seen from Eq. (\ref{eq4}) that the total absorption is the sum of contributions of each mode, which is the result of their opposite symmetry properties. 

Each of the two terms in Eq. (\ref{eq4}) will achieve a theoretical maximum of $50\%$ when the radiation rate of mode exactly match the dissipative loss rate of material at the resonant wavelength, $\omega=\omega_{j}$, $\gamma_{j}=\delta_{j}$, which is the classic critical coupling condition for single-mode, two-port system\cite{Wang2019, Xiao2020, Wang2020}. If the resonant frequency of the two modes falls apart, $|\omega_{1}-\omega_{2}|\gg\gamma_{j}$, the total absorption is mainly contributed by only one mode and is again limited to $50\%$. However, if the two modes are degenerate in frequency, $\omega=\omega_{1}=\omega_{2}$, and simultaneously the radiation rate equals the dissipative loss rate, $\gamma_{1}=\delta_{1}$, $\gamma_{2}=\delta_{2}$, the entire system reaches the so-called degenerate critical coupling state, and perfect absorption of $100\%$ will be achieved at this time. It must be pointed out that the dissipative loss rate is an intrinsic property of the absorbing material, which puts forward an essential constraint $\delta_{1}=\delta_{2}=\delta$ at the same wavelength. In combination with the above analysis, the requirements for the degenerate critical coupling are quite strict, i.e., $\omega=\omega_{1}=\omega_{2}$, and $\gamma_{1}=\gamma_{2}=\delta$, which have been excessively relaxed in some of previous works. 

\section{\label{sec3}Numerical results and discussions}

\begin{figure*}[htbp]
	\centering
	\includegraphics% Here is how to import EPS art
	[scale=0.60]{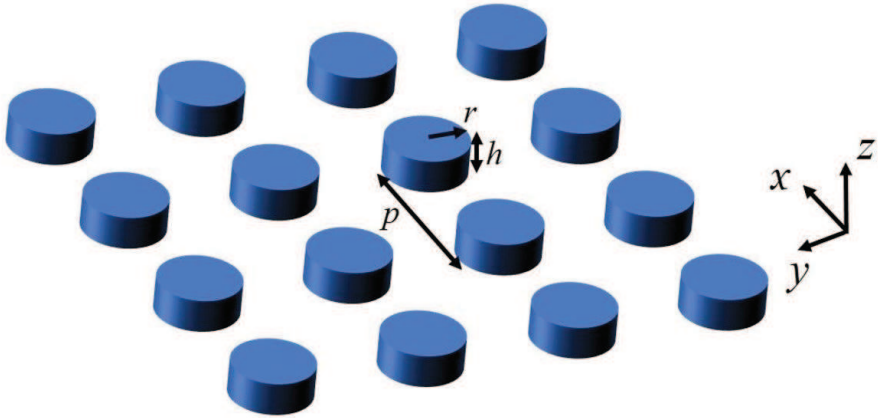}
	\caption{\label{fig1} Schematic of the proposed ultrathin semiconductor metasurface absorber comprising of the free-standing GaAs nanocylinders with the radius $r$, the height $h$, and the periodicity $p$.}
\end{figure*}

As an illustration of the theoretical concept of degenerate critical coupling presented above, we consider a two-port system of the free-standing GaAs metasurface absorber, which comprises of a GaAs nanocylinder in each unit cell, as shown in Fig. \ref{fig1}. The free-standing GaAs nanocylinders show the mirror symmetry about the $x$-$y$ plane lying in the center of the nanocylinders. The proposed structure is designed so that it supports the two resonance modes including an electric dipole and a magnetic dipole, when the structure is illuminated by the $x$-polarized incident plane waves propagating along the –$z$ axis. We utilize the finite-difference time-domain (FDTD) method to numerically simulate the optical responses of the proposed structure. In simulations, the GaAs nanocylinders have periodicity $p=650$ nm, the radius $r$ and the height $h$, and the wavelength-dependent parameters of the GaAs material (Palik data) is used\cite{Palik1998}. The spectral range in the simulation is from 700 nm to 1100 nm in the near infrared.

\begin{figure}[htbp]
	\centering
	\includegraphics% Here is how to import EPS art
	[scale=0.50]{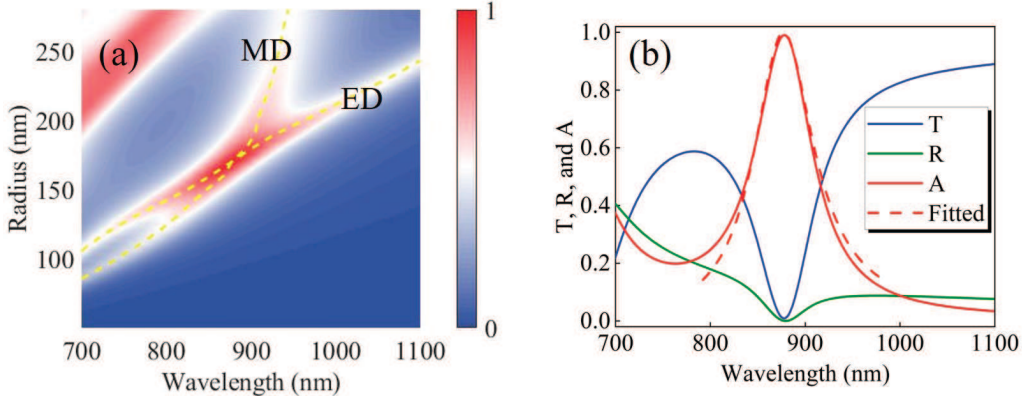}
	\caption{\label{fig2} (a) The simulated absorption spectra of the proposed structure as a function of wavelength and radius of the GaAs nanocylinders, with the nanocylinder height $h=140$ nm. The electric dipole and magnetic dipole modes are marked by dashed lines. (b) The simulated transmission, reflection and absorption spectra at the wavelength 878.236 nm of crossing resonance, and the theoretical absorption spectrum is denoted by the dashed line. }
\end{figure}

To achieve the perfect absorption in the proposed structure, we conduct a search in the three-dimensional space of parameters including the wavelength $\lambda$, the radius $r$, and the height $h$ of the GaAs nanocylinders. Fig. \ref{fig2}(a) shows the simulated absorption of the structure as a function of wavelength and the radius of the GaAs nanocylinders with a fixed height $h= 140$ nm, while the electric dipole and magnetic dipole modes are marked by dashed lines. As the radius increases, it is observed that the absorption of the structure for each single mode never exceeds $50\%$, and the absorption is considerably enhanced when the two modes start to overlap. Once the two resonance modes cross at the wavelength of 878.236 nm near the bandgap edge of GaAs where the intrinsic absorption of GaAs approaches zero, the perfect absorption is achieved at this point. When the radius continues to increase, leaving the degenerate critical coupling position, the absorption shows a downward trend. In Fig. \ref{fig2}(b), the peak absorption of $99.067\%$ occurs at 878.236 nm for the nanocylinder radius $r= 170$ nm and the height $h=140$ nm, while both transmission and reflection are significantly suppressed near this wavelength of the mode crossing. Based on the TCMT, the theoretical absorption spectrum agrees excellently with the simulated absorption spectrum in the vicinity of resonance. By fitting the absorption spectrum using Eq. (\ref{eq4}), the radiation rate and dissipative loss rate of the structure are obtained as $\gamma_{1}=\gamma_{2}=\delta_{1}=\delta_{2}$=47.04 THz at the resonance wavelength. Hence the condition of degenerate critical coupling is exactly fulfilled in this case, accounting for the perfect absorption of the structure.

\begin{figure}[htbp]
	\centering
	\includegraphics% Here is how to import EPS art
	[scale=0.50]{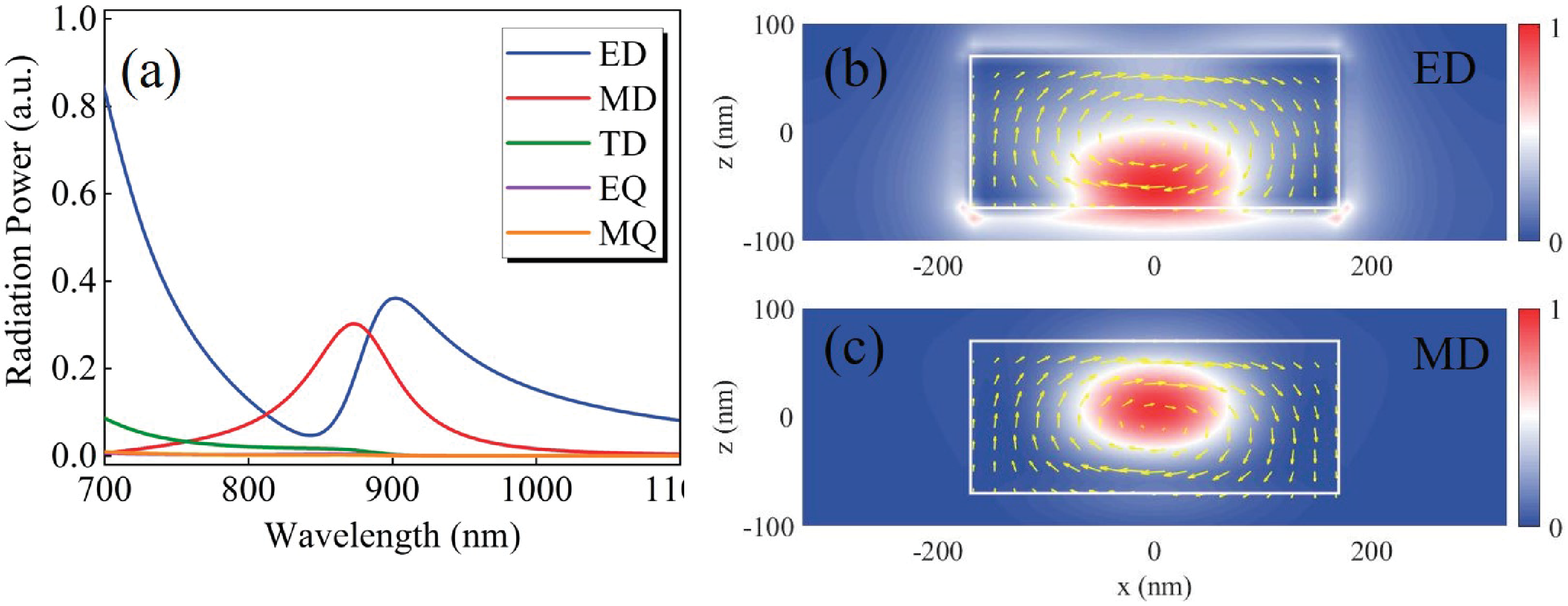}
	\caption{\label{fig3} (a) The multipole decomposition of the scattering cross sections, including the multipolar contributions from electric dipole (ED), magnetic dipole (MD), toroidal dipole (TD), electric quadrupole (EQ), and magnetic quadrupole (MQ). (b) The corresponding electric and magnetic field distributions in the $x$-$z$ plane at the resonance wavelength of 878.236 nm, and the arrows indicate the displacement currents. }
\end{figure}

To further unveil the contributions of the electric dipole and magnetic dipole modes to the perfect absorption under the degenerate coupling condition, the multipole decompositions of the scattering cross sections of the GaAs arrays are conducted, where the multipolar contributions from electric dipole, magnetic dipole, toroidal dipole, electric quadrupole, and magnetic quadrupole are shown in Fig. \ref{fig3}(a). The electric dipole and magnetic dipole modes dominate at 878.236 nm and spectrally oscillate in phase with each other, while the contributions of higher-order resonances including toroidal dipole, electric quadrupole, and magnetic quadrupole can be neglected within this wavelength region. According to the corresponding field distributions of the resonances in the $x$-$z$ plane inside each unit cell in Fig. \ref{fig3}(b), the electric dipole resonance is excited along the $x$ direction, and the excitation of the magnetic dipole resonance is in the $y$ direction and forms displacement current loop in the $x$-$z$ plane, satisfying the opposite symmetry requirement of the TCMT. Through the degenerate critical coupling of the two modes with opposite symmetry, the absorption of the structure is significantly enhanced at the wavelength of mode crossing.

\begin{figure}[htbp]
	\centering
	\includegraphics% Here is how to import EPS art
	[scale=0.50]{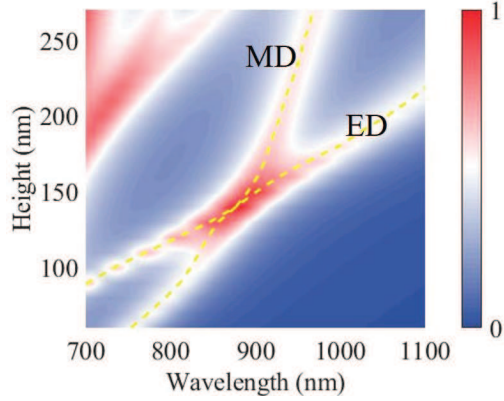}
	\caption{\label{fig4} The absorption spectra of the proposed structure as a function of wavelength and the height of the GaAs nanocylinders, with the nanocylinder radius $r=170$ nm. The electric dipole and magnetic dipole modes are marked by dashed lines.}
\end{figure}

When searching for the optimal geometry for the perfect absorption, the absorption spectra as a function of the GaAs nanocylinder height $h$ under normal incidence are also illustrated in Fig. \ref{fig4}, in addition to the dependence of absorption on the nanocylinder radius $r$ presented above. Similarly with the case in Fig. \ref{fig2}, the electric dipole and magnetic dipole modes of opposite symmetry are simultaneously excited within the wavelength of interest. It can be clearly observed that as the height $h$ increases with fixed $r=170$ nm, the electric dipole and magnetic dipole modes experience wavelength redshifts, leading to the change from separation to crossing and at last separation again, accompanied with the different levels of absorption enhancement. At the point of the mode crossing of 878.236 nm and $h=140$ nm, the conditions for degenerate critical coupling can be fulfilled, giving rise to the perfect absorption of the structure. Thus the geometry optimization and in-plane symmetry control of the absorption structures provide one simple and general way to achieve and fine-tune the degenerate condition for perfect absorption.

\begin{figure}[htbp]
	\centering
	\includegraphics% Here is how to import EPS art
	[scale=0.50]{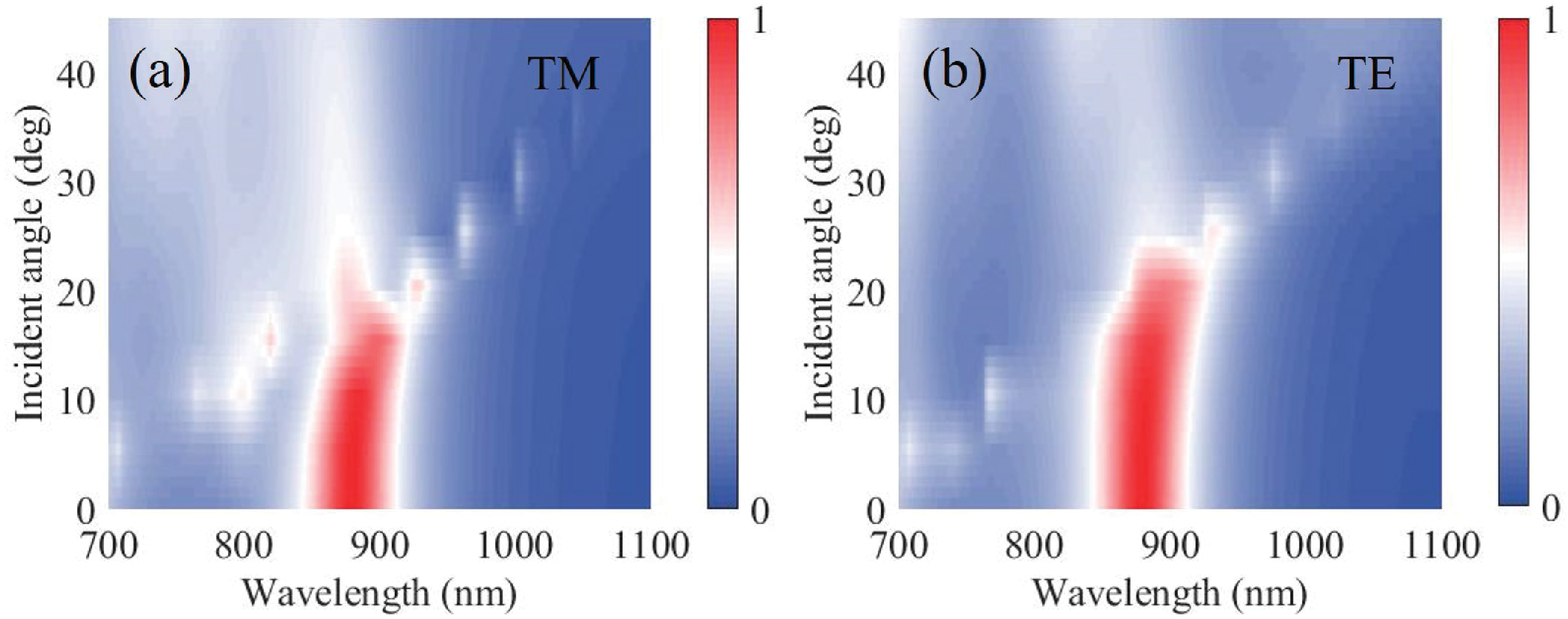}
	\caption{\label{fig5} The absorption spectra of the proposed structure with the GaAs nanocylinder radius $r=170$ nm and height $h=140$ nm under (a) TM and (b) TE polarization.}
\end{figure}

The dependence of the absorption performance in the proposed structure on the incident polarization and angle is also investigated. Fig. \ref{fig5} illustrates the simulated absorption spectra as a function of incident angle under TM and TE polarizations when the nanocylinder radius and height are fixed as $r= 170$ nm and $h=140$ nm. In comparison of Fig. \ref{fig5}(a) and \ref{fig5}(b), it can be seen that the absorption spectra are almost insensitive to the polarizations in the vicinity of wavelength 878.236 nm, showing high tolerance to the polarizations. Especially, the high absorption arising from the crossing electric dipole and magnetic dipole modes near this wavelength can be maintained within the polarization angle range of $25^{\circ}$.

\section{\label{sec4}Conclusions}

In conclusion, a kind of ultrathin semiconductor metasurfaces in the near infrared is theoretically and numerically demonstrated for perfect absorption based on the concept of degenerate critical coupling. In the two-port system comprising of the free-standing GaAs nanocylinders, the simultaneous excitation of the electric dipole and magnetic dipole modes with opposite symmetry leads to the enhanced absorption. By the geometry optimization such as adjusting the radius and height of the GaAs nanocylinder, the electric dipole and magnetic dipole modes reach their respective critical coupling state at the same resonance wavelength. At the point of mode crossing, the strict condition for degenerate critical coupling can be fulfilled and the perfect absorption is achieved in the exemplary structure, breaking the limit of $50\%$ absorption in the free-standing film. The TCMT and the multipole decompositions are also conducted for theoretical analysis, verifying the mechanism of the degenerate critical coupling. In addition, the proposed structure also shows the polarization-independent and angle-insensitive properties. This work provides the guidance for the ultrathin perfect absorption semiconductor metasurfaces for breaking the absorption limitation of free-standing films and the various limitations of metal solutions, and holds great potential for high-efficiency optical and optoelectronic devices in the future application.

\begin{acknowledgments}	
This work is supported by the National Natural Science Foundation of China (Grants No. 11847132, No. 11947065, No. 61901164, and No. 12004084), the Natural Science Foundation of Jiangxi Province (Grant No. 20202BAB211007), the Interdisciplinary Innovation Fund of Nanchang University (Grant No. 2019-9166-27060003), the Natural Science Research Project of Guizhou Minzu University (Grant No. GZMU[2019]YB22), and the China Scholarship Council (Grant No. 202008420045). The authors would also like to thank Dr. S. Li for her guidance on the effective multipole expansion.

W.C. and X.W. contributed equally to this work.
\end{acknowledgments}


%

%\bibliography{Ref-3.26}% Produces the bibliography via BibTeX.

\begin{thebibliography}{36}%
	\makeatletter
	\providecommand \@ifxundefined [1]{%
		\@ifx{#1\undefined}
	}%
	\providecommand \@ifnum [1]{%
		\ifnum #1\expandafter \@firstoftwo
		\else \expandafter \@secondoftwo
		\fi
	}%
	\providecommand \@ifx [1]{%
		\ifx #1\expandafter \@firstoftwo
		\else \expandafter \@secondoftwo
		\fi
	}%
	\providecommand \natexlab [1]{#1}%
	\providecommand \enquote  [1]{``#1''}%
	\providecommand \bibnamefont  [1]{#1}%
	\providecommand \bibfnamefont [1]{#1}%
	\providecommand \citenamefont [1]{#1}%
	\providecommand \href@noop [0]{\@secondoftwo}%
	\providecommand \href [0]{\begingroup \@sanitize@url \@href}%
	\providecommand \@href[1]{\@@startlink{#1}\@@href}%
	\providecommand \@@href[1]{\endgroup#1\@@endlink}%
	\providecommand \@sanitize@url [0]{\catcode `\\12\catcode `\$12\catcode
		`\&12\catcode `\#12\catcode `\^12\catcode `\_12\catcode `\%12\relax}%
	\providecommand \@@startlink[1]{}%
	\providecommand \@@endlink[0]{}%
	\providecommand \url  [0]{\begingroup\@sanitize@url \@url }%
	\providecommand \@url [1]{\endgroup\@href {#1}{\urlprefix }}%
	\providecommand \urlprefix  [0]{URL }%
	\providecommand \Eprint [0]{\href }%
	\providecommand \doibase [0]{http://dx.doi.org/}%
	\providecommand \selectlanguage [0]{\@gobble}%
	\providecommand \bibinfo  [0]{\@secondoftwo}%
	\providecommand \bibfield  [0]{\@secondoftwo}%
	\providecommand \translation [1]{[#1]}%
	\providecommand \BibitemOpen [0]{}%
	\providecommand \bibitemStop [0]{}%
	\providecommand \bibitemNoStop [0]{.\EOS\space}%
	\providecommand \EOS [0]{\spacefactor3000\relax}%
	\providecommand \BibitemShut  [1]{\csname bibitem#1\endcsname}%
	\let\auto@bib@innerbib\@empty
	%</preamble>
	\bibitem [{\citenamefont {Zhou}\ \emph {et~al.}(2016)\citenamefont {Zhou},
		\citenamefont {Tan}, \citenamefont {Wang}, \citenamefont {Xu}, \citenamefont
		{Yuan}, \citenamefont {Cai}, \citenamefont {Zhu},\ and\ \citenamefont
		{Zhu}}]{Zhou2016}%
	\BibitemOpen
	\bibfield  {author} {\bibinfo {author} {\bibfnamefont {L.}~\bibnamefont
			{Zhou}}, \bibinfo {author} {\bibfnamefont {Y.}~\bibnamefont {Tan}}, \bibinfo
		{author} {\bibfnamefont {J.}~\bibnamefont {Wang}}, \bibinfo {author}
		{\bibfnamefont {W.}~\bibnamefont {Xu}}, \bibinfo {author} {\bibfnamefont
			{Y.}~\bibnamefont {Yuan}}, \bibinfo {author} {\bibfnamefont {W.}~\bibnamefont
			{Cai}}, \bibinfo {author} {\bibfnamefont {S.}~\bibnamefont {Zhu}}, \ and\
		\bibinfo {author} {\bibfnamefont {J.}~\bibnamefont {Zhu}},\ }\href {\doibase
		10.1038/nphoton.2016.75} {\bibfield  {journal} {\bibinfo  {journal} {Nat.
				Photonics}\ }\textbf {\bibinfo {volume} {10}},\ \bibinfo {pages} {393}
		(\bibinfo {year} {2016})}\BibitemShut {NoStop}%
	\bibitem [{\citenamefont {Ma}\ \emph {et~al.}(2018)\citenamefont {Ma},
		\citenamefont {Yan}, \citenamefont {Huang}, \citenamefont {Wang},\ and\
		\citenamefont {Yang}}]{Ma2018}%
	\BibitemOpen
	\bibfield  {author} {\bibinfo {author} {\bibfnamefont {C.}~\bibnamefont
			{Ma}}, \bibinfo {author} {\bibfnamefont {J.}~\bibnamefont {Yan}}, \bibinfo
		{author} {\bibfnamefont {Y.}~\bibnamefont {Huang}}, \bibinfo {author}
		{\bibfnamefont {C.}~\bibnamefont {Wang}}, \ and\ \bibinfo {author}
		{\bibfnamefont {G.}~\bibnamefont {Yang}},\ }\href {\doibase
		10.1126/sciadv.aas9894} {\bibfield  {journal} {\bibinfo  {journal} {Sci.
				Adv.}\ }\textbf {\bibinfo {volume} {4}},\ \bibinfo {pages} {eaas9894}
		(\bibinfo {year} {2018})}\BibitemShut {NoStop}%
	\bibitem [{\citenamefont {Wang}\ \emph
		{et~al.}(2020{\natexlab{a}})\citenamefont {Wang}, \citenamefont {Yang},
		\citenamefont {Qi}, \citenamefont {Zhang},\ and\ \citenamefont
		{Cheng}}]{Wang2020a}%
	\BibitemOpen
	\bibfield  {author} {\bibinfo {author} {\bibfnamefont {Z.}~\bibnamefont
			{Wang}}, \bibinfo {author} {\bibfnamefont {P.}~\bibnamefont {Yang}}, \bibinfo
		{author} {\bibfnamefont {G.}~\bibnamefont {Qi}}, \bibinfo {author}
		{\bibfnamefont {Z.~M.}\ \bibnamefont {Zhang}}, \ and\ \bibinfo {author}
		{\bibfnamefont {P.}~\bibnamefont {Cheng}},\ }\href {\doibase
		10.1063/5.0005700} {\bibfield  {journal} {\bibinfo  {journal} {J. Appl.
				Phys.}\ }\textbf {\bibinfo {volume} {127}},\ \bibinfo {pages} {233102}
		(\bibinfo {year} {2020}{\natexlab{a}})}\BibitemShut {NoStop}%
	\bibitem [{\citenamefont {Tagliabue}\ \emph {et~al.}(2018)\citenamefont
		{Tagliabue}, \citenamefont {Jermyn}, \citenamefont {Sundararaman},
		\citenamefont {Welch}, \citenamefont {DuChene}, \citenamefont {Pala},
		\citenamefont {Davoyan}, \citenamefont {Narang},\ and\ \citenamefont
		{Atwater}}]{Tagliabue2018}%
	\BibitemOpen
	\bibfield  {author} {\bibinfo {author} {\bibfnamefont {G.}~\bibnamefont
			{Tagliabue}}, \bibinfo {author} {\bibfnamefont {A.~S.}\ \bibnamefont
			{Jermyn}}, \bibinfo {author} {\bibfnamefont {R.}~\bibnamefont
			{Sundararaman}}, \bibinfo {author} {\bibfnamefont {A.~J.}\ \bibnamefont
			{Welch}}, \bibinfo {author} {\bibfnamefont {J.~S.}\ \bibnamefont {DuChene}},
		\bibinfo {author} {\bibfnamefont {R.}~\bibnamefont {Pala}}, \bibinfo {author}
		{\bibfnamefont {A.~R.}\ \bibnamefont {Davoyan}}, \bibinfo {author}
		{\bibfnamefont {P.}~\bibnamefont {Narang}}, \ and\ \bibinfo {author}
		{\bibfnamefont {H.~A.}\ \bibnamefont {Atwater}},\ }\href {\doibase
		10.1038/s41467-018-05968-x} {\bibfield  {journal} {\bibinfo  {journal} {Nat.
				Commun.}\ }\textbf {\bibinfo {volume} {9}},\ \bibinfo {pages} {3394}
		(\bibinfo {year} {2018})}\BibitemShut {NoStop}%
	\bibitem [{\citenamefont {Guo}\ \emph {et~al.}(2020)\citenamefont {Guo},
		\citenamefont {Li}, \citenamefont {Liu}, \citenamefont {Yin}, \citenamefont
		{Wang}, \citenamefont {Ni}, \citenamefont {Fu}, \citenamefont {Yu},
		\citenamefont {Xu}, \citenamefont {Shi}, \citenamefont {Ma}, \citenamefont
		{Gao}, \citenamefont {Tong},\ and\ \citenamefont {Dai}}]{Guo2020}%
	\BibitemOpen
	\bibfield  {author} {\bibinfo {author} {\bibfnamefont {J.}~\bibnamefont
			{Guo}}, \bibinfo {author} {\bibfnamefont {J.}~\bibnamefont {Li}}, \bibinfo
		{author} {\bibfnamefont {C.}~\bibnamefont {Liu}}, \bibinfo {author}
		{\bibfnamefont {Y.}~\bibnamefont {Yin}}, \bibinfo {author} {\bibfnamefont
			{W.}~\bibnamefont {Wang}}, \bibinfo {author} {\bibfnamefont {Z.}~\bibnamefont
			{Ni}}, \bibinfo {author} {\bibfnamefont {Z.}~\bibnamefont {Fu}}, \bibinfo
		{author} {\bibfnamefont {H.}~\bibnamefont {Yu}}, \bibinfo {author}
		{\bibfnamefont {Y.}~\bibnamefont {Xu}}, \bibinfo {author} {\bibfnamefont
			{Y.}~\bibnamefont {Shi}}, \bibinfo {author} {\bibfnamefont {Y.}~\bibnamefont
			{Ma}}, \bibinfo {author} {\bibfnamefont {S.}~\bibnamefont {Gao}}, \bibinfo
		{author} {\bibfnamefont {L.}~\bibnamefont {Tong}}, \ and\ \bibinfo {author}
		{\bibfnamefont {D.}~\bibnamefont {Dai}},\ }\href {\doibase
		10.1038/s41377-020-0263-6} {\bibfield  {journal} {\bibinfo  {journal} {Light
				Sci. Appl.}\ }\textbf {\bibinfo {volume} {9}},\ \bibinfo {pages} {29}
		(\bibinfo {year} {2020})}\BibitemShut {NoStop}%
	\bibitem [{\citenamefont {Rodrigo}\ \emph {et~al.}(2015)\citenamefont
		{Rodrigo}, \citenamefont {Limaj}, \citenamefont {Janner}, \citenamefont
		{Etezadi}, \citenamefont {de~Abajo}, \citenamefont {Pruneri},\ and\
		\citenamefont {Altug}}]{Rodrigo2015}%
	\BibitemOpen
	\bibfield  {author} {\bibinfo {author} {\bibfnamefont {D.}~\bibnamefont
			{Rodrigo}}, \bibinfo {author} {\bibfnamefont {O.}~\bibnamefont {Limaj}},
		\bibinfo {author} {\bibfnamefont {D.}~\bibnamefont {Janner}}, \bibinfo
		{author} {\bibfnamefont {D.}~\bibnamefont {Etezadi}}, \bibinfo {author}
		{\bibfnamefont {F.~J.~G.}\ \bibnamefont {de~Abajo}}, \bibinfo {author}
		{\bibfnamefont {V.}~\bibnamefont {Pruneri}}, \ and\ \bibinfo {author}
		{\bibfnamefont {H.}~\bibnamefont {Altug}},\ }\href {\doibase
		10.1126/science.aab2051} {\bibfield  {journal} {\bibinfo  {journal}
			{Science}\ }\textbf {\bibinfo {volume} {349}},\ \bibinfo {pages} {165}
		(\bibinfo {year} {2015})}\BibitemShut {NoStop}%
	\bibitem [{\citenamefont {Tan}\ \emph {et~al.}(2018)\citenamefont {Tan},
		\citenamefont {Yan}, \citenamefont {Wang}, \citenamefont {Zhou},\ and\
		\citenamefont {Hou}}]{Tan2018}%
	\BibitemOpen
	\bibfield  {author} {\bibinfo {author} {\bibfnamefont {S.}~\bibnamefont
			{Tan}}, \bibinfo {author} {\bibfnamefont {F.}~\bibnamefont {Yan}}, \bibinfo
		{author} {\bibfnamefont {W.}~\bibnamefont {Wang}}, \bibinfo {author}
		{\bibfnamefont {H.}~\bibnamefont {Zhou}}, \ and\ \bibinfo {author}
		{\bibfnamefont {Y.}~\bibnamefont {Hou}},\ }\href {\doibase
		10.1088/2040-8986/aab66e} {\bibfield  {journal} {\bibinfo  {journal} {J.
				Opt.}\ }\textbf {\bibinfo {volume} {20}},\ \bibinfo {pages} {055101}
		(\bibinfo {year} {2018})}\BibitemShut {NoStop}%
	\bibitem [{\citenamefont {Ren}\ \emph {et~al.}(2020)\citenamefont {Ren},
		\citenamefont {Lin}, \citenamefont {Zhi},\ and\ \citenamefont
		{Li}}]{Ren2020}%
	\BibitemOpen
	\bibfield  {author} {\bibinfo {author} {\bibfnamefont {Z.}~\bibnamefont
			{Ren}}, \bibinfo {author} {\bibfnamefont {Z.}~\bibnamefont {Lin}}, \bibinfo
		{author} {\bibfnamefont {X.}~\bibnamefont {Zhi}}, \ and\ \bibinfo {author}
		{\bibfnamefont {M.}~\bibnamefont {Li}},\ }\href {\doibase
		10.1016/j.optmat.2019.109575} {\bibfield  {journal} {\bibinfo  {journal}
			{Opt. Mater.}\ }\textbf {\bibinfo {volume} {99}},\ \bibinfo {pages} {109575}
		(\bibinfo {year} {2020})}\BibitemShut {NoStop}%
	\bibitem [{\citenamefont {Zhao}\ \emph {et~al.}(2020)\citenamefont {Zhao},
		\citenamefont {Zhang}, \citenamefont {Yang}, \citenamefont {Li},
		\citenamefont {Feng}, \citenamefont {Quan}, \citenamefont {Yang},\ and\
		\citenamefont {Xiao}}]{Zhao2020}%
	\BibitemOpen
	\bibfield  {author} {\bibinfo {author} {\bibfnamefont {W.}~\bibnamefont
			{Zhao}}, \bibinfo {author} {\bibfnamefont {Y.}~\bibnamefont {Zhang}},
		\bibinfo {author} {\bibfnamefont {J.}~\bibnamefont {Yang}}, \bibinfo {author}
		{\bibfnamefont {J.}~\bibnamefont {Li}}, \bibinfo {author} {\bibfnamefont
			{Y.}~\bibnamefont {Feng}}, \bibinfo {author} {\bibfnamefont {M.}~\bibnamefont
			{Quan}}, \bibinfo {author} {\bibfnamefont {Z.}~\bibnamefont {Yang}}, \ and\
		\bibinfo {author} {\bibfnamefont {S.}~\bibnamefont {Xiao}},\ }\href {\doibase
		10.1039/d0nr02972f} {\bibfield  {journal} {\bibinfo  {journal} {Nanoscale}\
		}\textbf {\bibinfo {volume} {12}},\ \bibinfo {pages} {18056} (\bibinfo {year}
		{2020})}\BibitemShut {NoStop}%
	\bibitem [{\citenamefont {Kim}\ \emph {et~al.}(2016)\citenamefont {Kim},
		\citenamefont {Park}, \citenamefont {Esfandyarpour}, \citenamefont {Pecora},
		\citenamefont {Kik},\ and\ \citenamefont {Brongersma}}]{Kim2016}%
	\BibitemOpen
	\bibfield  {author} {\bibinfo {author} {\bibfnamefont {S.~J.}\ \bibnamefont
			{Kim}}, \bibinfo {author} {\bibfnamefont {J.}~\bibnamefont {Park}}, \bibinfo
		{author} {\bibfnamefont {M.}~\bibnamefont {Esfandyarpour}}, \bibinfo {author}
		{\bibfnamefont {E.~F.}\ \bibnamefont {Pecora}}, \bibinfo {author}
		{\bibfnamefont {P.~G.}\ \bibnamefont {Kik}}, \ and\ \bibinfo {author}
		{\bibfnamefont {M.~L.}\ \bibnamefont {Brongersma}},\ }\href {\doibase
		10.1021/acs.nanolett.6b01198} {\bibfield  {journal} {\bibinfo  {journal}
			{Nano Lett.}\ }\textbf {\bibinfo {volume} {16}},\ \bibinfo {pages} {3801}
		(\bibinfo {year} {2016})}\BibitemShut {NoStop}%
	\bibitem [{\citenamefont {Dallenbach}\ and\ \citenamefont
		{Kleinsteuber}(1938)}]{Dallenbach1938}%
	\BibitemOpen
	\bibfield  {author} {\bibinfo {author} {\bibfnamefont {W.}~\bibnamefont
			{Dallenbach}}\ and\ \bibinfo {author} {\bibfnamefont {W.}~\bibnamefont
			{Kleinsteuber}},\ }\href@noop {} {\bibfield  {journal} {\bibinfo  {journal}
			{Hochfreq. Elektroak}\ }\textbf {\bibinfo {volume} {51}},\ \bibinfo {pages}
		{152} (\bibinfo {year} {1938})}\BibitemShut {NoStop}%
	\bibitem [{\citenamefont {Salisbury}(1952)}]{Salisbury1952}%
	\BibitemOpen
	\bibfield  {author} {\bibinfo {author} {\bibfnamefont {W.~W.}\ \bibnamefont
			{Salisbury}},\ }\href@noop {} {\bibfield  {journal} {\bibinfo  {journal} {US
				Patent}\ ,\ \bibinfo {pages} {2,599,944}} (\bibinfo {year}
		{1952})}\BibitemShut {NoStop}%
	\bibitem [{\citenamefont {Landy}\ \emph {et~al.}(2008)\citenamefont {Landy},
		\citenamefont {Sajuyigbe}, \citenamefont {Mock}, \citenamefont {Smith},\ and\
		\citenamefont {Padilla}}]{Landy2008}%
	\BibitemOpen
	\bibfield  {author} {\bibinfo {author} {\bibfnamefont {N.~I.}\ \bibnamefont
			{Landy}}, \bibinfo {author} {\bibfnamefont {S.}~\bibnamefont {Sajuyigbe}},
		\bibinfo {author} {\bibfnamefont {J.~J.}\ \bibnamefont {Mock}}, \bibinfo
		{author} {\bibfnamefont {D.~R.}\ \bibnamefont {Smith}}, \ and\ \bibinfo
		{author} {\bibfnamefont {W.~J.}\ \bibnamefont {Padilla}},\ }\href {\doibase
		10.1103/physrevlett.100.207402} {\bibfield  {journal} {\bibinfo  {journal}
			{Phys. Rev. Lett.}\ }\textbf {\bibinfo {volume} {100}},\ \bibinfo {pages}
		{207402} (\bibinfo {year} {2008})}\BibitemShut {NoStop}%
	\bibitem [{\citenamefont {Liu}\ \emph {et~al.}(2010)\citenamefont {Liu},
		\citenamefont {Mesch}, \citenamefont {Weiss}, \citenamefont {Hentschel},\
		and\ \citenamefont {Giessen}}]{Liu2010}%
	\BibitemOpen
	\bibfield  {author} {\bibinfo {author} {\bibfnamefont {N.}~\bibnamefont
			{Liu}}, \bibinfo {author} {\bibfnamefont {M.}~\bibnamefont {Mesch}}, \bibinfo
		{author} {\bibfnamefont {T.}~\bibnamefont {Weiss}}, \bibinfo {author}
		{\bibfnamefont {M.}~\bibnamefont {Hentschel}}, \ and\ \bibinfo {author}
		{\bibfnamefont {H.}~\bibnamefont {Giessen}},\ }\href {\doibase
		10.1021/nl9041033} {\bibfield  {journal} {\bibinfo  {journal} {Nano Lett.}\
		}\textbf {\bibinfo {volume} {10}},\ \bibinfo {pages} {2342} (\bibinfo {year}
		{2010})}\BibitemShut {NoStop}%
	\bibitem [{\citenamefont {Xiong}\ \emph {et~al.}(2017)\citenamefont {Xiong},
		\citenamefont {Zhang}, \citenamefont {Zhu}, \citenamefont {Yuan},\ and\
		\citenamefont {Qin}}]{Xiong2017}%
	\BibitemOpen
	\bibfield  {author} {\bibinfo {author} {\bibfnamefont {F.}~\bibnamefont
			{Xiong}}, \bibinfo {author} {\bibfnamefont {J.}~\bibnamefont {Zhang}},
		\bibinfo {author} {\bibfnamefont {Z.}~\bibnamefont {Zhu}}, \bibinfo {author}
		{\bibfnamefont {X.}~\bibnamefont {Yuan}}, \ and\ \bibinfo {author}
		{\bibfnamefont {S.}~\bibnamefont {Qin}},\ }\href {\doibase
		10.1088/2040-8986/aa7292} {\bibfield  {journal} {\bibinfo  {journal} {J.
				Opt.}\ }\textbf {\bibinfo {volume} {19}},\ \bibinfo {pages} {075002}
		(\bibinfo {year} {2017})}\BibitemShut {NoStop}%
	\bibitem [{\citenamefont {Luo}\ \emph {et~al.}(2018)\citenamefont {Luo},
		\citenamefont {Liu}, \citenamefont {Cheng}, \citenamefont {Liu},
		\citenamefont {Lin},\ and\ \citenamefont {Wang}}]{Luo2018}%
	\BibitemOpen
	\bibfield  {author} {\bibinfo {author} {\bibfnamefont {X.}~\bibnamefont
			{Luo}}, \bibinfo {author} {\bibfnamefont {Z.}~\bibnamefont {Liu}}, \bibinfo
		{author} {\bibfnamefont {Z.}~\bibnamefont {Cheng}}, \bibinfo {author}
		{\bibfnamefont {J.}~\bibnamefont {Liu}}, \bibinfo {author} {\bibfnamefont
			{Q.}~\bibnamefont {Lin}}, \ and\ \bibinfo {author} {\bibfnamefont
			{L.}~\bibnamefont {Wang}},\ }\href {\doibase 10.1364/oe.26.033918} {\bibfield
		{journal} {\bibinfo  {journal} {Opt. Express}\ }\textbf {\bibinfo {volume}
			{26}},\ \bibinfo {pages} {33918} (\bibinfo {year} {2018})}\BibitemShut
	{NoStop}%
	\bibitem [{\citenamefont {Cheng}\ \emph {et~al.}(2020)\citenamefont {Cheng},
		\citenamefont {Zhao},\ and\ \citenamefont {Li}}]{Cheng2020}%
	\BibitemOpen
	\bibfield  {author} {\bibinfo {author} {\bibfnamefont {Y.}~\bibnamefont
			{Cheng}}, \bibinfo {author} {\bibfnamefont {H.}~\bibnamefont {Zhao}}, \ and\
		\bibinfo {author} {\bibfnamefont {C.}~\bibnamefont {Li}},\ }\href {\doibase
		10.1016/j.optmat.2020.110369} {\bibfield  {journal} {\bibinfo  {journal}
			{Opt. Mater.}\ }\textbf {\bibinfo {volume} {109}},\ \bibinfo {pages} {110369}
		(\bibinfo {year} {2020})}\BibitemShut {NoStop}%
	\bibitem [{\citenamefont {Alaee}\ \emph {et~al.}(2017)\citenamefont {Alaee},
		\citenamefont {Albooyeh},\ and\ \citenamefont {Rockstuhl}}]{Alaee2017}%
	\BibitemOpen
	\bibfield  {author} {\bibinfo {author} {\bibfnamefont {R.}~\bibnamefont
			{Alaee}}, \bibinfo {author} {\bibfnamefont {M.}~\bibnamefont {Albooyeh}}, \
		and\ \bibinfo {author} {\bibfnamefont {C.}~\bibnamefont {Rockstuhl}},\ }\href
	{\doibase 10.1088/1361-6463/aa94a8} {\bibfield  {journal} {\bibinfo
			{journal} {J. Phys. D: Appl. Phys.}\ }\textbf {\bibinfo {volume} {50}},\
		\bibinfo {pages} {503002} (\bibinfo {year} {2017})}\BibitemShut {NoStop}%
	\bibitem [{\citenamefont {Kuznetsov}\ \emph {et~al.}(2016)\citenamefont
		{Kuznetsov}, \citenamefont {Miroshnichenko}, \citenamefont {Brongersma},
		\citenamefont {Kivshar},\ and\ \citenamefont {Luk'yanchuk}}]{Kuznetsov2016}%
	\BibitemOpen
	\bibfield  {author} {\bibinfo {author} {\bibfnamefont {A.~I.}\ \bibnamefont
			{Kuznetsov}}, \bibinfo {author} {\bibfnamefont {A.~E.}\ \bibnamefont
			{Miroshnichenko}}, \bibinfo {author} {\bibfnamefont {M.~L.}\ \bibnamefont
			{Brongersma}}, \bibinfo {author} {\bibfnamefont {Y.~S.}\ \bibnamefont
			{Kivshar}}, \ and\ \bibinfo {author} {\bibfnamefont {B.}~\bibnamefont
			{Luk'yanchuk}},\ }\href {\doibase 10.1126/science.aag2472} {\bibfield
		{journal} {\bibinfo  {journal} {Science}\ }\textbf {\bibinfo {volume}
			{354}},\ \bibinfo {pages} {aag2472} (\bibinfo {year} {2016})}\BibitemShut
	{NoStop}%
	\bibitem [{\citenamefont {Jahani}\ and\ \citenamefont
		{Jacob}(2016)}]{Jahani2016}%
	\BibitemOpen
	\bibfield  {author} {\bibinfo {author} {\bibfnamefont {S.}~\bibnamefont
			{Jahani}}\ and\ \bibinfo {author} {\bibfnamefont {Z.}~\bibnamefont {Jacob}},\
	}\href {\doibase 10.1038/nnano.2015.304} {\bibfield  {journal} {\bibinfo
			{journal} {Nat. Nanotechnol.}\ }\textbf {\bibinfo {volume} {11}},\ \bibinfo
		{pages} {23} (\bibinfo {year} {2016})}\BibitemShut {NoStop}%
	\bibitem [{\citenamefont {Decker}\ and\ \citenamefont
		{Staude}(2016)}]{Decker2016}%
	\BibitemOpen
	\bibfield  {author} {\bibinfo {author} {\bibfnamefont {M.}~\bibnamefont
			{Decker}}\ and\ \bibinfo {author} {\bibfnamefont {I.}~\bibnamefont
			{Staude}},\ }\href {\doibase 10.1088/2040-8978/18/10/103001} {\bibfield
		{journal} {\bibinfo  {journal} {J. Opt.}\ }\textbf {\bibinfo {volume} {18}},\
		\bibinfo {pages} {103001} (\bibinfo {year} {2016})}\BibitemShut {NoStop}%
	\bibitem [{\citenamefont {Koshelev}\ \emph {et~al.}(2018)\citenamefont
		{Koshelev}, \citenamefont {Lepeshov}, \citenamefont {Liu}, \citenamefont
		{Bogdanov},\ and\ \citenamefont {Kivshar}}]{Koshelev2018}%
	\BibitemOpen
	\bibfield  {author} {\bibinfo {author} {\bibfnamefont {K.}~\bibnamefont
			{Koshelev}}, \bibinfo {author} {\bibfnamefont {S.}~\bibnamefont {Lepeshov}},
		\bibinfo {author} {\bibfnamefont {M.}~\bibnamefont {Liu}}, \bibinfo {author}
		{\bibfnamefont {A.}~\bibnamefont {Bogdanov}}, \ and\ \bibinfo {author}
		{\bibfnamefont {Y.}~\bibnamefont {Kivshar}},\ }\href {\doibase
		10.1103/physrevlett.121.193903} {\bibfield  {journal} {\bibinfo  {journal}
			{Phys. Rev. Lett.}\ }\textbf {\bibinfo {volume} {121}},\ \bibinfo {pages}
		{193903} (\bibinfo {year} {2018})}\BibitemShut {NoStop}%
	\bibitem [{\citenamefont {Li}\ \emph {et~al.}(2019)\citenamefont {Li},
		\citenamefont {Zhou}, \citenamefont {Liu},\ and\ \citenamefont
		{Xiao}}]{Li2019}%
	\BibitemOpen
	\bibfield  {author} {\bibinfo {author} {\bibfnamefont {S.}~\bibnamefont
			{Li}}, \bibinfo {author} {\bibfnamefont {C.}~\bibnamefont {Zhou}}, \bibinfo
		{author} {\bibfnamefont {T.}~\bibnamefont {Liu}}, \ and\ \bibinfo {author}
		{\bibfnamefont {S.}~\bibnamefont {Xiao}},\ }\href {\doibase
		10.1103/physreva.100.063803} {\bibfield  {journal} {\bibinfo  {journal}
			{Phys. Rev. A}\ }\textbf {\bibinfo {volume} {100}},\ \bibinfo {pages}
		{063803} (\bibinfo {year} {2019})}\BibitemShut {NoStop}%
	\bibitem [{\citenamefont {Huang}\ \emph {et~al.}(2021)\citenamefont {Huang},
		\citenamefont {Xu}, \citenamefont {Rahmani}, \citenamefont {Neshev},\ and\
		\citenamefont {Miroshnichenko}}]{Huang2021}%
	\BibitemOpen
	\bibfield  {author} {\bibinfo {author} {\bibfnamefont {L.}~\bibnamefont
			{Huang}}, \bibinfo {author} {\bibfnamefont {L.}~\bibnamefont {Xu}}, \bibinfo
		{author} {\bibfnamefont {M.}~\bibnamefont {Rahmani}}, \bibinfo {author}
		{\bibfnamefont {D.}~\bibnamefont {Neshev}}, \ and\ \bibinfo {author}
		{\bibfnamefont {A.~E.}\ \bibnamefont {Miroshnichenko}},\ }\href {\doibase
		10.1117/1.ap.3.1.016004} {\bibfield  {journal} {\bibinfo  {journal} {Adv.
				Photonics}\ }\textbf {\bibinfo {volume} {3}},\ \bibinfo {pages} {016004}
		(\bibinfo {year} {2021})}\BibitemShut {NoStop}%
	\bibitem [{\citenamefont {Piper}\ \emph {et~al.}(2014)\citenamefont {Piper},
		\citenamefont {Liu},\ and\ \citenamefont {Fan}}]{Piper2014}%
	\BibitemOpen
	\bibfield  {author} {\bibinfo {author} {\bibfnamefont {J.~R.}\ \bibnamefont
			{Piper}}, \bibinfo {author} {\bibfnamefont {V.}~\bibnamefont {Liu}}, \ and\
		\bibinfo {author} {\bibfnamefont {S.}~\bibnamefont {Fan}},\ }\href {\doibase
		10.1063/1.4885517} {\bibfield  {journal} {\bibinfo  {journal} {Appl. Phys.
				Lett.}\ }\textbf {\bibinfo {volume} {104}},\ \bibinfo {pages} {251110}
		(\bibinfo {year} {2014})}\BibitemShut {NoStop}%
	\bibitem [{\citenamefont {Ming}\ \emph {et~al.}(2017)\citenamefont {Ming},
		\citenamefont {Liu}, \citenamefont {Sun},\ and\ \citenamefont
		{Padilla}}]{Ming2017}%
	\BibitemOpen
	\bibfield  {author} {\bibinfo {author} {\bibfnamefont {X.}~\bibnamefont
			{Ming}}, \bibinfo {author} {\bibfnamefont {X.}~\bibnamefont {Liu}}, \bibinfo
		{author} {\bibfnamefont {L.}~\bibnamefont {Sun}}, \ and\ \bibinfo {author}
		{\bibfnamefont {W.~J.}\ \bibnamefont {Padilla}},\ }\href {\doibase
		10.1364/oe.25.024658} {\bibfield  {journal} {\bibinfo  {journal} {Opt.
				Express}\ }\textbf {\bibinfo {volume} {25}},\ \bibinfo {pages} {24658}
		(\bibinfo {year} {2017})}\BibitemShut {NoStop}%
	\bibitem [{\citenamefont {Tian}\ \emph {et~al.}(2018)\citenamefont {Tian},
		\citenamefont {Luo}, \citenamefont {Li}, \citenamefont {Pei}, \citenamefont
		{Du},\ and\ \citenamefont {Qiu}}]{Tian2018}%
	\BibitemOpen
	\bibfield  {author} {\bibinfo {author} {\bibfnamefont {J.}~\bibnamefont
			{Tian}}, \bibinfo {author} {\bibfnamefont {H.}~\bibnamefont {Luo}}, \bibinfo
		{author} {\bibfnamefont {Q.}~\bibnamefont {Li}}, \bibinfo {author}
		{\bibfnamefont {X.}~\bibnamefont {Pei}}, \bibinfo {author} {\bibfnamefont
			{K.}~\bibnamefont {Du}}, \ and\ \bibinfo {author} {\bibfnamefont
			{M.}~\bibnamefont {Qiu}},\ }\href {\doibase 10.1002/lpor.201800076}
	{\bibfield  {journal} {\bibinfo  {journal} {Laser Photonics Rev.}\ }\textbf
		{\bibinfo {volume} {12}},\ \bibinfo {pages} {1800076} (\bibinfo {year}
		{2018})}\BibitemShut {NoStop}%
	\bibitem [{\citenamefont {Tian}\ \emph {et~al.}(2020)\citenamefont {Tian},
		\citenamefont {Li}, \citenamefont {Belov}, \citenamefont {Sinha},
		\citenamefont {Qian},\ and\ \citenamefont {Qiu}}]{Tian2020}%
	\BibitemOpen
	\bibfield  {author} {\bibinfo {author} {\bibfnamefont {J.}~\bibnamefont
			{Tian}}, \bibinfo {author} {\bibfnamefont {Q.}~\bibnamefont {Li}}, \bibinfo
		{author} {\bibfnamefont {P.~A.}\ \bibnamefont {Belov}}, \bibinfo {author}
		{\bibfnamefont {R.~K.}\ \bibnamefont {Sinha}}, \bibinfo {author}
		{\bibfnamefont {W.}~\bibnamefont {Qian}}, \ and\ \bibinfo {author}
		{\bibfnamefont {M.}~\bibnamefont {Qiu}},\ }\href {\doibase
		10.1021/acsphotonics.0c00003} {\bibfield  {journal} {\bibinfo  {journal}
			{{ACS} Photonics}\ }\textbf {\bibinfo {volume} {7}},\ \bibinfo {pages} {1436}
		(\bibinfo {year} {2020})}\BibitemShut {NoStop}%
	\bibitem [{\citenamefont {Mitrofanov}\ \emph {et~al.}(2020)\citenamefont
		{Mitrofanov}, \citenamefont {Hale}, \citenamefont {Vabishchevich},
		\citenamefont {Luk}, \citenamefont {Addamane}, \citenamefont {Reno},\ and\
		\citenamefont {Brener}}]{Mitrofanov2020}%
	\BibitemOpen
	\bibfield  {author} {\bibinfo {author} {\bibfnamefont {O.}~\bibnamefont
			{Mitrofanov}}, \bibinfo {author} {\bibfnamefont {L.~L.}\ \bibnamefont
			{Hale}}, \bibinfo {author} {\bibfnamefont {P.~P.}\ \bibnamefont
			{Vabishchevich}}, \bibinfo {author} {\bibfnamefont {T.~S.}\ \bibnamefont
			{Luk}}, \bibinfo {author} {\bibfnamefont {S.~J.}\ \bibnamefont {Addamane}},
		\bibinfo {author} {\bibfnamefont {J.~L.}\ \bibnamefont {Reno}}, \ and\
		\bibinfo {author} {\bibfnamefont {I.}~\bibnamefont {Brener}},\ }\href
	{\doibase 10.1063/5.0019883} {\bibfield  {journal} {\bibinfo  {journal}
			{{APL} Photonics}\ }\textbf {\bibinfo {volume} {5}},\ \bibinfo {pages}
		{101304} (\bibinfo {year} {2020})}\BibitemShut {NoStop}%
	\bibitem [{\citenamefont {Hale}\ \emph {et~al.}(2020)\citenamefont {Hale},
		\citenamefont {Vabischevich}, \citenamefont {Siday}, \citenamefont {Harris},
		\citenamefont {Luk}, \citenamefont {Addamane}, \citenamefont {Reno},
		\citenamefont {Brener},\ and\ \citenamefont {Mitrofanov}}]{Hale2020}%
	\BibitemOpen
	\bibfield  {author} {\bibinfo {author} {\bibfnamefont {L.~L.}\ \bibnamefont
			{Hale}}, \bibinfo {author} {\bibfnamefont {P.~P.}\ \bibnamefont
			{Vabischevich}}, \bibinfo {author} {\bibfnamefont {T.}~\bibnamefont {Siday}},
		\bibinfo {author} {\bibfnamefont {C.~T.}\ \bibnamefont {Harris}}, \bibinfo
		{author} {\bibfnamefont {T.~S.}\ \bibnamefont {Luk}}, \bibinfo {author}
		{\bibfnamefont {S.~J.}\ \bibnamefont {Addamane}}, \bibinfo {author}
		{\bibfnamefont {J.~L.}\ \bibnamefont {Reno}}, \bibinfo {author}
		{\bibfnamefont {I.}~\bibnamefont {Brener}}, \ and\ \bibinfo {author}
		{\bibfnamefont {O.}~\bibnamefont {Mitrofanov}},\ }\href {\doibase
		10.1364/oe.404249} {\bibfield  {journal} {\bibinfo  {journal} {Opt. Express}\
		}\textbf {\bibinfo {volume} {28}},\ \bibinfo {pages} {35284} (\bibinfo {year}
		{2020})}\BibitemShut {NoStop}%
	\bibitem [{\citenamefont {Fan}\ \emph {et~al.}(2021)\citenamefont {Fan},
		\citenamefont {Shadrivov}, \citenamefont {Miroshnichenko},\ and\
		\citenamefont {Padilla}}]{Fan2021}%
	\BibitemOpen
	\bibfield  {author} {\bibinfo {author} {\bibfnamefont {K.}~\bibnamefont
			{Fan}}, \bibinfo {author} {\bibfnamefont {I.~V.}\ \bibnamefont {Shadrivov}},
		\bibinfo {author} {\bibfnamefont {A.~E.}\ \bibnamefont {Miroshnichenko}}, \
		and\ \bibinfo {author} {\bibfnamefont {W.~J.}\ \bibnamefont {Padilla}},\
	}\href {\doibase 10.1364/oe.421187} {\bibfield  {journal} {\bibinfo
			{journal} {Opt. Express}\ }\textbf {\bibinfo {volume} {29}},\ \bibinfo
		{pages} {10518} (\bibinfo {year} {2021})}\BibitemShut {NoStop}%
	\bibitem [{\citenamefont {Suh}\ \emph {et~al.}(2004)\citenamefont {Suh},
		\citenamefont {Wang},\ and\ \citenamefont {Fan}}]{Suh2004}%
	\BibitemOpen
	\bibfield  {author} {\bibinfo {author} {\bibfnamefont {W.}~\bibnamefont
			{Suh}}, \bibinfo {author} {\bibfnamefont {Z.}~\bibnamefont {Wang}}, \ and\
		\bibinfo {author} {\bibfnamefont {S.}~\bibnamefont {Fan}},\ }\href {\doibase
		10.1109/jqe.2004.834773} {\bibfield  {journal} {\bibinfo  {journal} {{IEEE}
				J. Quantum Electron.}\ }\textbf {\bibinfo {volume} {40}},\ \bibinfo {pages}
		{1511} (\bibinfo {year} {2004})}\BibitemShut {NoStop}%
	\bibitem [{\citenamefont {Wang}\ \emph {et~al.}(2019)\citenamefont {Wang},
		\citenamefont {Chen}, \citenamefont {Zhang}, \citenamefont {Zeng},
		\citenamefont {Zhang}, \citenamefont {Liu}, \citenamefont {Shi},\ and\
		\citenamefont {Zi}}]{Wang2019}%
	\BibitemOpen
	\bibfield  {author} {\bibinfo {author} {\bibfnamefont {J.}~\bibnamefont
			{Wang}}, \bibinfo {author} {\bibfnamefont {A.}~\bibnamefont {Chen}}, \bibinfo
		{author} {\bibfnamefont {Y.}~\bibnamefont {Zhang}}, \bibinfo {author}
		{\bibfnamefont {J.}~\bibnamefont {Zeng}}, \bibinfo {author} {\bibfnamefont
			{Y.}~\bibnamefont {Zhang}}, \bibinfo {author} {\bibfnamefont
			{X.}~\bibnamefont {Liu}}, \bibinfo {author} {\bibfnamefont {L.}~\bibnamefont
			{Shi}}, \ and\ \bibinfo {author} {\bibfnamefont {J.}~\bibnamefont {Zi}},\
	}\href {\doibase 10.1103/physrevb.100.075407} {\bibfield  {journal} {\bibinfo
			{journal} {Phys. Rev. B}\ }\textbf {\bibinfo {volume} {100}},\ \bibinfo
		{pages} {075407} (\bibinfo {year} {2019})}\BibitemShut {NoStop}%
	\bibitem [{\citenamefont {Xiao}\ \emph {et~al.}(2020)\citenamefont {Xiao},
		\citenamefont {Liu}, \citenamefont {Wang}, \citenamefont {Liu},\ and\
		\citenamefont {Zhou}}]{Xiao2020}%
	\BibitemOpen
	\bibfield  {author} {\bibinfo {author} {\bibfnamefont {S.}~\bibnamefont
			{Xiao}}, \bibinfo {author} {\bibfnamefont {T.}~\bibnamefont {Liu}}, \bibinfo
		{author} {\bibfnamefont {X.}~\bibnamefont {Wang}}, \bibinfo {author}
		{\bibfnamefont {X.}~\bibnamefont {Liu}}, \ and\ \bibinfo {author}
		{\bibfnamefont {C.}~\bibnamefont {Zhou}},\ }\href {\doibase
		10.1103/physrevb.102.085410} {\bibfield  {journal} {\bibinfo  {journal}
			{Phys. Rev. B}\ }\textbf {\bibinfo {volume} {102}},\ \bibinfo {pages}
		{085410} (\bibinfo {year} {2020})}\BibitemShut {NoStop}%
	\bibitem [{\citenamefont {Wang}\ \emph
		{et~al.}(2020{\natexlab{b}})\citenamefont {Wang}, \citenamefont {Duan},
		\citenamefont {Chen}, \citenamefont {Zhou}, \citenamefont {Liu},\ and\
		\citenamefont {Xiao}}]{Wang2020}%
	\BibitemOpen
	\bibfield  {author} {\bibinfo {author} {\bibfnamefont {X.}~\bibnamefont
			{Wang}}, \bibinfo {author} {\bibfnamefont {J.}~\bibnamefont {Duan}}, \bibinfo
		{author} {\bibfnamefont {W.}~\bibnamefont {Chen}}, \bibinfo {author}
		{\bibfnamefont {C.}~\bibnamefont {Zhou}}, \bibinfo {author} {\bibfnamefont
			{T.}~\bibnamefont {Liu}}, \ and\ \bibinfo {author} {\bibfnamefont
			{S.}~\bibnamefont {Xiao}},\ }\href {\doibase 10.1103/physrevb.102.155432}
	{\bibfield  {journal} {\bibinfo  {journal} {Phys. Rev. B}\ }\textbf {\bibinfo
			{volume} {102}},\ \bibinfo {pages} {155432} (\bibinfo {year}
		{2020}{\natexlab{b}})}\BibitemShut {NoStop}%
	\bibitem [{\citenamefont {Palik}(1998)}]{Palik1998}%
	\BibitemOpen
	\bibfield  {author} {\bibinfo {author} {\bibfnamefont {E.~D.}\ \bibnamefont
			{Palik}},\ }\href@noop {} {\emph {\bibinfo {title} {Handbook of optical
				constants of solids}}}\ (\bibinfo  {publisher} {Academic},\ \bibinfo {year}
	{1998})\BibitemShut {NoStop}%
\end{thebibliography}

\end{document}